\documentclass[%
 aip,
 amsmath,amssymb,
 reprint,%
]{revtex4-1}

\usepackage{graphicx}
\usepackage{dcolumn}
\usepackage{bm}

\usepackage[utf8]{inputenc}
\usepackage[T1]{fontenc}
\usepackage{mathptmx}
\usepackage{mathtools}

\usepackage{hyperref}

\newcommand{\mydot}[1]{\overset{\boldsymbol{.}}{#1}}

\begin{document}

\preprint{AIP/123-QED}

\title[Solitary states in the mean-field limit]{Solitary states in the mean-field limit}

\author{N. Kruk}
	\affiliation{Technische Universit\"{a}t Darmstadt, Rundeturmstrasse 12, 64283, Darmstadt, Germany}
\author{Y. Maistrenko}
	\affiliation{Forschungszentrum J\"ulich GmbH, Wilhelm-Johnen-Stra\ss e, 52428 J\"ulich, Germany}
	\affiliation{National Academy of Sciences of Ukraine, Tereshchenkivska St. 3, 01601, Kyiv, Ukraine}
\author{H. Koeppl}
	\altaffiliation{Author to whom correspondence should be addressed}
	\email{heinz.koeppl@bcs.tu-darmstadt.de.}
	\affiliation{Technische Universit\"{a}t Darmstadt, Rundeturmstrasse 12, 64283, Darmstadt, Germany}


\date{\today}

\begin{abstract}
We study active matter systems where the orientational dynamics of underlying self-propelled particles obey second order equations. By primarily concentrating on a spatially homogeneous setup for particle distribution, our analysis combines theories of active matter and oscillatory networks. For such systems, we analyze the appearance of solitary states via a homoclinic bifurcation as a mechanism of the frequency clustering. By introducing noise, we establish a stochastic version of solitary states and derive the mean-field limit described by a partial differential equation for a one-particle probability density function, which one might call the continuum Kuramoto model with inertia and noise. By studying this limit, we establish second order phase transitions between polar order and disorder. The combination of both analytical and numerical approaches in our study demonstrates an excellent qualitative agreement between mean-field and finite size models.
\end{abstract}

\maketitle

\begin{quotation}
Self-organization in large collectives of interacting particles is a fascinating phenomenon that is not completely understood yet. We study how spatially homogeneous particle ensembles behave subject to second-order rules of motion. Spatial homogeneity allows us to simplify the description of particle dynamics to that of their orientations only. This leads us to the Kuramoto model with inertia and allows us to regard particles as network oscillators, for which solitary states that naturally arise in systems of coupled pendula and power grids have recently been discovered.
The goal of this study is to analyze the appearance of solitary states from the point of view of the active matter theory, particularly in the mean-field limit and under the influence of noise.
\end{quotation}

\section{Introduction}

Collective motion of large systems of agents is a fascinating phenomenon found in many living as well as artificial environments. Ample examples range from bacterial swarming through flocking of birds and schooling of fish to robotic ensembles, to name a few \cite{vicsek2012phys_rep,gompper2020jpcm}. The first prominent model to describe self-organized dynamics of such systems was the Vicsek model\cite{vicsek1995prl}. It postulates that each agent or particle changes its direction of motion in discrete time steps to the direction averaged across its neighborhood. Later, it has been shown \cite{degond2008mmmas} that under an appropriate scaling, the Vicsek model can be recast into a continuous time form, where the temporal update for the direction of motion is effectively the same as the Kuramoto model for networks of coupled oscillators \cite{kuramoto1984springer}. As a result, in situations where the spatial information about collective dynamics is negligible, the study of a particle model coincides with the study of an oscillator model. In the view of recent increase of attention to the Kuramoto model with inertia, we are interested in analyzing how a similar modification would influence self-propelled particle motion.

To be more precise, let us consider a system of $N$ particles moving in a two-dimensional space with periodic boundaries of size $L$ with constant velocity magnitude $v_0\in\mathbb{R}$. The state of a particle is given by a position $r_i\in\mathbb{U}^2,\mathbb{U}:=\mathbb{R}/(L\mathbb{Z})$, orientation $\varphi_i\in\mathbb{T}, \mathbb{T}:=\mathbb{R}/(2\pi\mathbb{Z})$, and angular velocity $\omega_i\in\mathbb{R}$.
We describe particles' motion with the following system of ordinary differential equations (ODEs):
\begin{equation}
\label{eq:particle_ode_nonlocal}
\begin{aligned}
	\mydot{r}_i(t) &= v_0e_i(t), \\
	\mydot{\varphi}_i(t) &= \omega_i(t), \\
	\mydot{\omega}_i(t) &= -\xi\omega_i(t) + \frac{\sigma}{\vert B_\varrho^i\vert} \sum_{j\in B_\varrho^i} \sin(\varphi_j(t) - \varphi_i(t) - \alpha),
\end{aligned}
\end{equation}
where $e_i=(\cos\varphi_i,\sin\varphi_i)$ denotes particle's orientation; $\xi\in\mathbb{R}_+$ is a rotational friction coefficient; $\sigma\in\mathbb{R}_+$ controls the strength of alignment within a neighborhood $B_\varrho^i := \left\{ j=1,\dots,N \mid \Vert r_i - r_j \Vert \leq \varrho \right\}$ of radius $\varrho\in\mathbb{U}$; $\alpha\in\mathbb{T}$ is a phase lag. By nondimensionalization, we find that $\frac{\xi B}{A} \gg 1$, where $A$ and $B$ are time and phase scales, respectively, defines the overdamped limit for orientational dynamics. In this limit, which implies $\mydot{\omega}_i\approx0$, with symmetric interaction potential, i.e., $\alpha=0$, Eq.~\ref{eq:particle_ode_nonlocal} becomes the known continuous time formulation of the Vicsek model \cite{degond2008mmmas}. We remark that it has been shown that an alternative second order model in phase proves relevant in explaining oscillations in bacterial swarming \cite{chen2017nature}.

\begin{figure}[]
	\centering
	\includegraphics[width=0.4\textwidth]{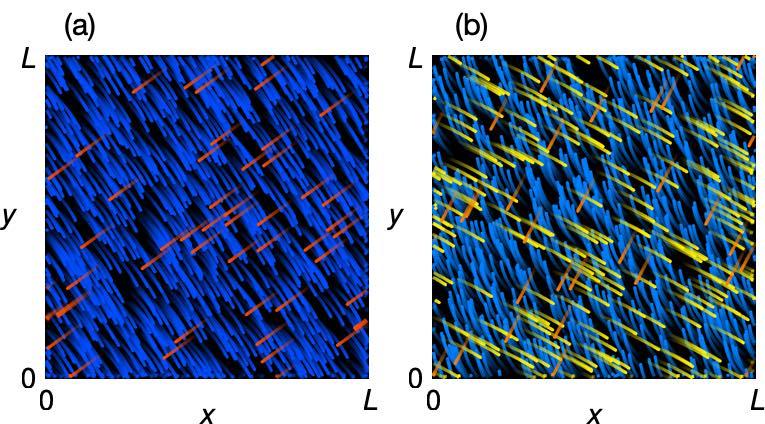}
	\caption{Examples of spatially homogeneous particle dynamics (see corresponding movies in the Supplementary Material (SM) and in \cite{bcs_youtube_channel}) generated by \eqref{eq:particle_ode_nonlocal} with $N=10^3$ where particles self-organize into (a) two and (b) three frequency groups. Particle color denotes instantaneous angular velocity. Background is shown in black. Particles are represented as stripes with transparency increasing backward in time. Other parameters are the same as in Fig.~\ref{fig:space_time_plots_sorted}(a) and (d), respectively.}
	\label{fig:spatially_homogeneous_particle_dynamics}
\end{figure}

Eq.~\eqref{eq:particle_ode_nonlocal} becomes the Kuramoto model with inertia in situations where the dynamics of position variables $r_i$ become negligible. Namely, in the context of active matter theory, this arises in the following situations. First, if we consider global interactions between particles, i.e., $\varrho \geq \frac{L}{2}$, spatial inhomogeneity of the last term in \eqref{eq:particle_ode_nonlocal} becomes irrelevant. Second, it is a common phenomenon for active matter systems that in the hydrodynamic limit $\varrho\rightarrow0_+$, particle dynamics may become spatially homogeneous. Therefore, as the first step towards understanding general dynamics of \eqref{eq:particle_ode_nonlocal}, we will restrict ourselves to its spatially homogeneous formulation. Examples of such particle motion, which results from second-order angular dynamics only, are presented in Fig.~\ref{fig:spatially_homogeneous_particle_dynamics}.

The Kuramoto model \cite{kuramoto1984springer} has gained a lot of attention in the last two decades due to the discovery of the striking coexistence of synchronized and desynchronized groups in networks of coupled oscillators, which became known as {\it chimera states} \cite{kuramoto2002npcs,abrams2004prl}. Since that, chimera states  where obtained in different fields,  see recent review papers \cite{panaggio2015nonlinearity,scholl2016epjst,omelchenko2019njp}, also in self-propelled systems \cite{kruk2018pre}. Subsequent studies on the Kuramoto model with inertia have revealed the appearance of yet another collective oscillatory motion termed {\it solitary states} \cite{maistrenko2014pre,jaros2015pre,jaros2018chaos}. They satisfy the definition of \textit{weak} chimera states \cite{ashwin2015chaos} but, nevertheless, reflect a qualitatively distinct situation where only a single oscillator or a relatively small group of oscillators splits off from the main synchronized cluster and starts to rotate with a different averaged frequency, i.e., Poincar\'e rotation number (cf. Fig.~\ref{fig:spatially_homogeneous_particle_dynamics} and movies in SM and in \cite{bcs_youtube_channel} for examples of such motion in a self-propelled particle context \eqref{eq:particle_ode_nonlocal}). An importance of this kind of behavior follows from the fact that solitary states naturally arise in realistic networks with inertia, such as coupled pendula \cite{kapitaniak2014sci_rep}, power grids \cite{taher2019pre,hellmann2020nat_comm}, and adaptive networks  \cite{berner2019arxiv} but,  they are not possible in the paradigmatic standard Kuramoto model without inertia.

\section{Solitary phenomena}

\begin{figure}[b]
	\centering
	\includegraphics[width=0.5\textwidth]{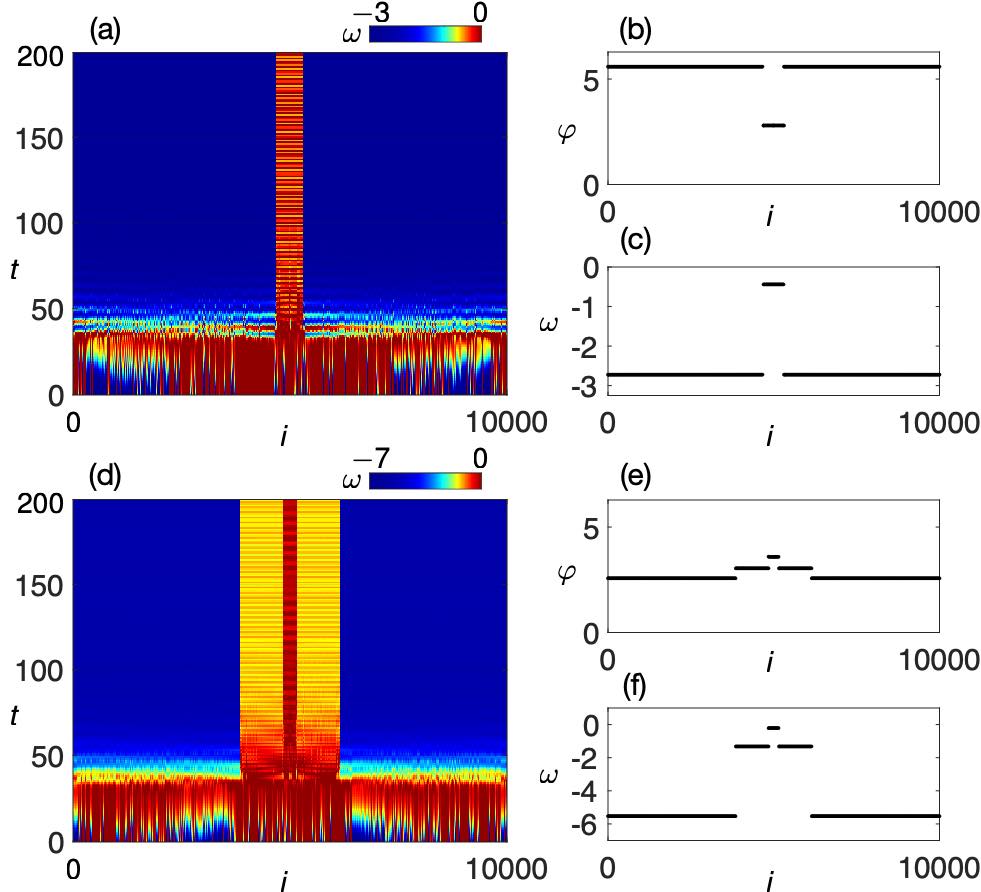}
	\caption{Temporal evolution towards a solitary state with (a) one frequency cluster ($\alpha=0.3$) and (d) two frequency clusters ($\alpha=0.8$) starting from random initial conditions as space-time plots. Color corresponds to instantaneous frequency $\omega$. (b,c) and (e,f) Respective profiles of instantaneous phase $\varphi$ and instantaneous frequency $\omega$. Frequency profiles correspond to respective horizontal intersections of space-time plots at $t=200$. Oscillators are reordered according to the value of instantaneous frequency at $t=200$ from (c) and (f), respectively. Other parameters are $\xi=0.1$, $\sigma=1$, $N=10^4$.}
	\label{fig:space_time_plots_sorted}
\end{figure}

Let us consider a spatially homogeneous formulation of particle dynamics \eqref{eq:particle_ode_nonlocal}, which we now refer to as oscillators. The state of each oscillator is given by a phase $\varphi_i\in\mathbb{T}$ and a frequency $\omega_i\in\mathbb{R}$. To study oscillatory dynamics without the influence of spatial distribution, we assume that oscillators are globally coupled and obey the following system of ODEs:
\begin{equation}
\label{eq:particle_ode_global}
\begin{aligned}
	\mydot{\varphi}_i(t) &= \omega_i(t), \\
	\mydot{\omega}_i(t) &= -\xi\omega_i(t) + \frac{\sigma}{N} \sum_{j=1}^N \sin(\varphi_j(t) - \varphi_i(t) - \alpha),
\end{aligned}
\end{equation}
where $\xi\in\mathbb{R}_+$ is a friction coefficient, $\sigma\in\mathbb{R}_+$ controls the strength of coupling, and $\alpha\in\mathbb{T}$ is a phase lag. The last term in \eqref{eq:particle_ode_global} favors synchronization between oscillators. The presence of the phase lag $\alpha$ induces additional rotation of the oscillators with respect to the average orientation of all neighbors. Eq.~\eqref{eq:particle_ode_global} contains three parameters $\xi$, $\sigma$, and $\alpha$, one of which can be eliminated by appropriate time scaling. For convenience, we put $\xi=0.1$ \cite{jaros2018chaos} for the rest of the paper and consider $\alpha\geq0$.

Numerical investigation of Eq.~\ref{eq:particle_ode_global} reveals the following. First, in the absence of the phase lag, i.e., when $\alpha=0$, all oscillators are stationary and completely synchronized. For small $\alpha>0$, the oscillators remain synchronized but rotating with the angular frequency $\omega^* = -\frac{\sigma}{\xi}\sin\alpha$.
Upon a further increase of $\alpha$ (cf. Fig.~\ref{fig:solitary_cascade}(a)) (and as long as $\xi < 2\sqrt{\sigma\cos\alpha}$, see Appendix~B and subsequent discussion), a group of oscillators split off from the majority and begins to rotate with a separate frequency. This type of dynamics has been termed as a \textit{solitary state} \cite{jaros2018chaos}.

The formation of solitary states starting from random initial conditions is illustrated in Fig.~\ref{fig:space_time_plots_sorted}(a)). At the beginning, the oscillators are disordered each pointing in its own direction and with its own frequency. Soon after, they gradually synchronize with respect to both phase $\varphi$ and frequency $\omega$ (cf. Fig.~\ref{fig:space_time_plots_sorted}(a), $t\lessapprox50$). At this point, one can already observe the formation of the second group of oscillators that are not in sync with the majority.  Subsequently, the division between two groups becomes more pronounced and the variations of $\varphi$ and $\omega$ inside each of them tend to minimize (cf. Fig.~\ref{fig:space_time_plots_sorted}(b,c)). 

\begin{figure}[t]
	\centering
	\includegraphics[width=0.5\textwidth]{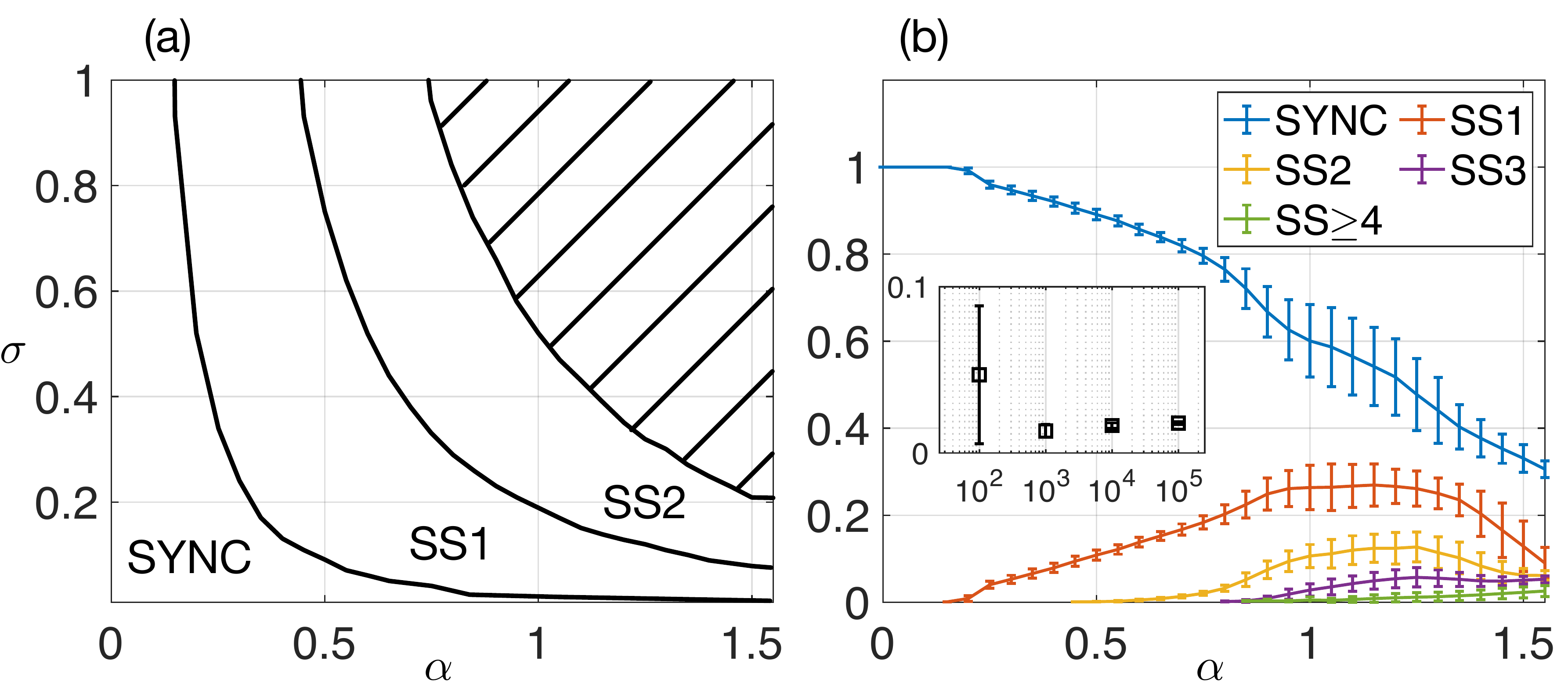}
	\caption{(a) Phase diagram of oscillatory dynamics due to \eqref{eq:particle_ode_global}. The synchronized motion (SYNC) is stable in all of the displayed domain. SS1 and SS2 denote regions where one and two groups of solitary oscillators emerge, respectively. Oblique hatching denotes a region where three and more solitary groups as well as chimeras appear. The phase diagram has been obtained with a continuation method for $N=10^3$. (b) Fraction of oscillators from different groups versus the phase lag $\alpha$ out of $200$ experiments per parameter tuple. Error bar indicates standard deviation with respect to different initial configurations. The inset shows the fraction of solitary oscillators in SS1 versus different population sizes ($\alpha=0.3$). Other parameters are $\xi=0.1$, $\sigma=1$.}
	\label{fig:solitary_cascade}
\end{figure}

Our simulations confirm that at further increase of the coupling strength $\sigma$ or the phase lag $\alpha$, the second solitary cluster emerges (cf. Fig.~\ref{fig:space_time_plots_sorted}(d-f) and Fig.~\ref{fig:solitary_cascade}(a)). The temporal evolution of a system towards such a state starting from random initial conditions is initially the same as described above for the solitary state with the only one frequency cluster. As it can be seen from Fig.~\ref{fig:space_time_plots_sorted}(d), during the initial stage, the first solitary cluster develops and soon after, at some point ($t\approx50$), the second smaller solitary group starts to rotate with their own distinctive frequencies. As time advances, this latter group synchronizes and rotates with a frequency different from frequencies of the other two clusters (cf. Fig.~\ref{fig:space_time_plots_sorted}(e,f)). Continuing in this way, we observe a cascade of solitary states with an increasing number of frequency clusters (cf. Fig.~\ref{fig:solitary_cascade}(b)). For $N=10^3$, we observe up to four additional such frequency clusters. Upon the further increase of the phase lag ($\alpha\gtrapprox1.3$ for $\sigma=1$), all solitary clusters merge and their averaged frequencies are continuously distributed over some range. In this paper, we primarily concentrate on the solitary states with one solitary group and will not discuss the rest of the cascade.

The number of solitary oscillators in frequency clusters depends essentially on initial conditions (cf. the inset in Fig.~\ref{fig:solitary_cascade}(b)). For small population sizes $N\sim10^2$, this dependence is strong. However, upon the increase of the system size $N$, these fluctuations decrease and the total fraction of oscillators in a solitary group tends to a certain limit. One can see that for population sizes $N\sim10^3$ and larger, the variance in the size drops drastically.

It has been shown \cite{jaros2018chaos} that the mechanism for one solitary oscillator to emerge is a homoclinic bifurcation of a saddle at some $\alpha=\alpha_1(\sigma)$. After the bifurcation, oscillators are separated into two populations, each with their own phase and frequency $(\varphi_0,\omega_0)$ and $(\varphi_1,\omega_1)$, respectively. System's dynamics evolves on a two-dimensional cylinder $\Omega^2$, $\Omega := \mathbb{T}\times\mathbb{R}$. It can easily be reduced to a one-dimensional cylinder for the difference variables $\Delta\varphi=\varphi_1-\varphi_0$ and $\Delta\omega=\omega_1-\omega_0$. Analyzing the dynamics of these difference variables (see the details in Appendix~B), we find two equilibria: a sink $O=(0,0)$ and a saddle  $S=(\pi-2\beta,0)$ with $\beta=\arctan[(1-2/N)\tan\alpha]$. The origin $O$ corresponds to complete synchronization and is a stable focus for $\xi<2\sqrt{\sigma\cos\alpha}$ (otherwise, it is a stable node). The second equilibrium $S$ is a saddle.
At $\alpha=\alpha_1$, a homoclinic orbit $\gamma_1$ is created in the moment when the unstable manifold of $S$ spans the phase space and comes back to $S$ as its stable manifold. It signifies the appearance of a solitary oscillator in \eqref{eq:particle_ode_global}. For $\alpha>\alpha_1$ the phase portrait contains a stable limit cycle that coexists with the stable focus $O$. Note that if the coupling strength $\sigma$ is kept constant and the friction coefficient is considerably increased, i.e, $\xi>2\sqrt{\sigma\cos\alpha}$, $O$ becomes a stable node and solitary oscillators do not emerge.

As the next step, we want to understand how two solitary oscillators appear given the aforementioned mechanism for the emergence of one such oscillator. Suppose that the system consists of $N-2$ synchronized oscillators and two solitary ones. The fraction of each of the solitary oscillators equals $w=1/N$ of the whole population. Let $(\varphi_0,\omega_0)$ denote phase and frequency of each synchronized oscillator and $(\varphi_1,\omega_1)$ and $(\varphi_2,\omega_2)$ denote the same variables of two solitary oscillators. In terms of difference variables $\Delta\varphi_{1,2}=\varphi_{1,2}-\varphi_0$ and $\Delta\omega_{1,2}=\omega_{1,2}-\omega_0$, system's dynamics are completely governed by (see Appendix~C)
\begin{equation}
\label{eq:ode_for_two_differences_symmetric}
\begin{aligned}
	\Delta\mydot{\varphi}_1 &= \Delta\omega_1, \\
	\Delta\mydot{\omega}_1 &= -\xi\Delta\omega_1 - \sigma R \sin(\Delta\varphi_1 + \beta) + \sigma B \\
	&+ \sigma w \sin(\Delta\varphi_2 - \Delta\varphi_1 - \alpha) - \sigma w \sin(\Delta\varphi_2 - \alpha), \\
	\Delta\mydot{\varphi}_2 &= \Delta\omega_2, \\
	\Delta\mydot{\omega}_2 &= -\xi\Delta\omega_2 - \sigma R \sin(\Delta\varphi_2 + \beta) + \sigma B \\
	&+ \sigma w \sin(\Delta\varphi_1 - \Delta\varphi_2 - \alpha) - \sigma w \sin(\Delta\varphi_1 - \alpha),
\end{aligned}
\end{equation}
where $A = (1-w) \cos\alpha$, $B = (1-3w) \sin\alpha$, $R = \sqrt{A^2 + B^2}$, and $\beta = \arctan(B/A)$. This system defines a flow on a two-dimensional cylinder $\Omega^2$. Note that Eqs.~\eqref{eq:ode_for_two_differences_symmetric} are symmetric with respect to the diagonal plane
\begin{equation}
\label{eq:diagonal_plane}
	D := \left\{ (\Delta\varphi_1,\Delta\omega_1,\Delta\varphi_2,\Delta\omega_2) \in \Omega^2 \mid  \Delta\varphi_1=\Delta\varphi_2, \Delta\omega_1=\Delta\omega_2 \right\}.
\end{equation}

\begin{figure}[]
	\centering
	\includegraphics[width=0.5\textwidth]{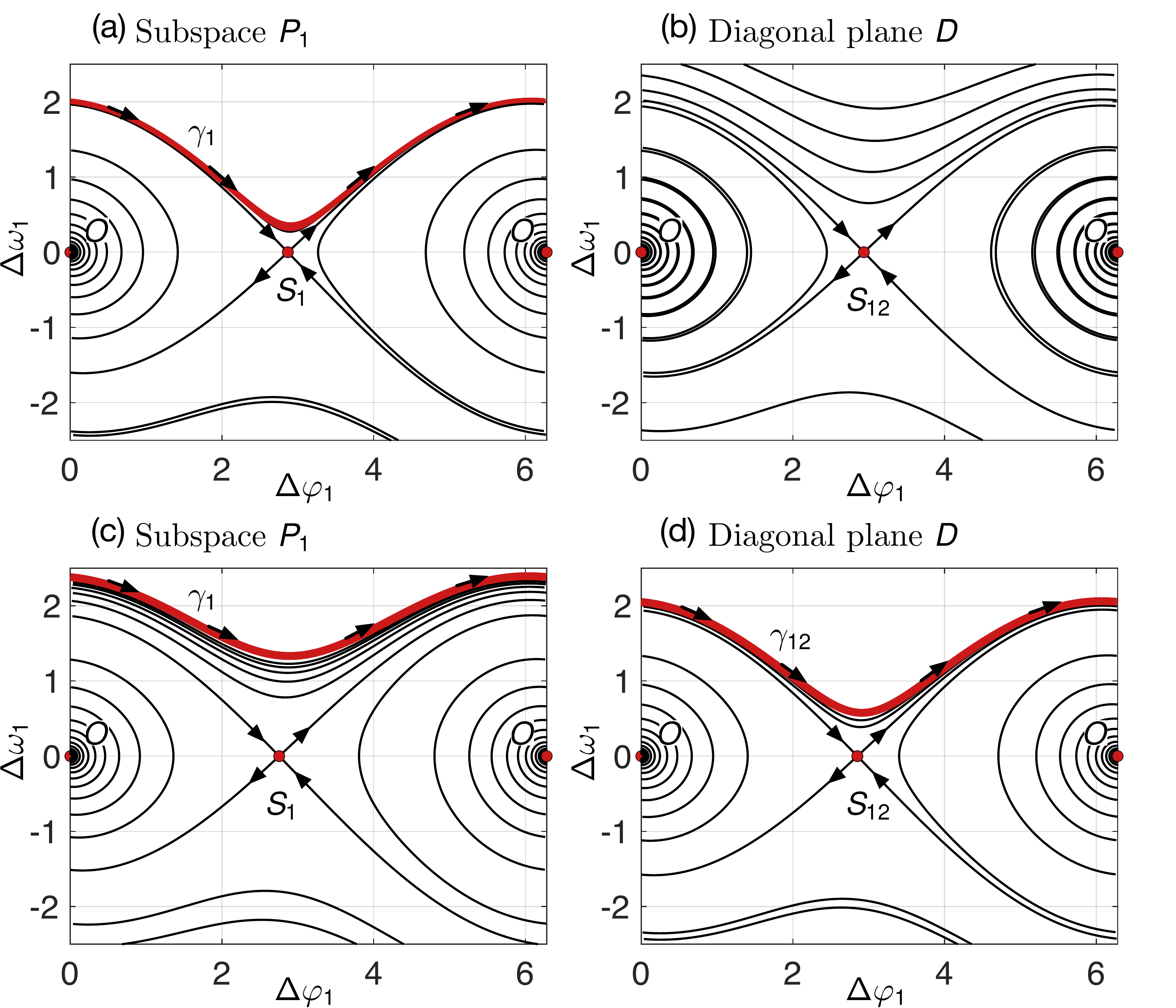}
	\caption{Subspaces of phase portraits of \eqref{eq:ode_for_two_differences_symmetric} for different values of the phase lag $\alpha$. (a) The plane $P_1$ \eqref{eq:subspace_of_first_solitary_particle} and (b) the diagonal plane $D$ \eqref{eq:diagonal_plane} of the first solitary oscillator, respectively, after appearance of a periodic orbit $\gamma_1 \subset P_1$, i.e., $\alpha_1<\alpha<\alpha_2$. (c) The plane $P_1$ and (d) the diagonal plane $D$ of the first solitary oscillator, respectively, after appearance of a periodic orbit $\gamma_{12} \subset D$, i.e., $\alpha>\alpha_2$. Black arrows around saddle points indicate their respective stable and unstable manifolds. Black arrows along periodic orbits indicate the direction of motion along them.}
	\label{fig:projected_phase_portrait_for_two_solitary_particles}
\end{figure}

The system \eqref{eq:ode_for_two_differences_symmetric} has four equilibria $O = (0,0,0,0)$, $S_1 = \left(\pi-2\beta',0,0,0\right)$, $S_2 = \left(0,0,\pi-2\beta',0\right)$, and $S_{12} = \left(\pi-2\beta'',0,\pi-2\beta'',0\right)$, where we have denoted $A' = \cos\alpha$, $B' = (1-2w) \sin\alpha$, $R' = \sqrt{A'^2 + B'^2}$, and $\beta' = \arctan(B'/A')$; $A'' = \cos\alpha$, $B'' = (1-4w) \sin\alpha$, $R'' = \sqrt{A''^2 + B''^2}$, and $\beta'' = \arctan(B'' / A'')$. $O$ is a stable focus as long as $\xi<2\sqrt{\sigma\cos\alpha}$; otherwise, it is a stable node. It corresponds to the synchronization of both oscillators with the main group. The fixed points $S_1$ and $S_2$ are of a saddle type. Two of their eigenvalues $\lambda_1>0$ and $\lambda_2<0$ are the same as in the previously discussed one-dimensional case; two remaining eigenvalues are equal $\lambda_{3,4} = \frac{1}{2} (-\xi \pm \sqrt{\xi^2 - 4 \sigma \cos\alpha (1-2w) / R'^2})$. Since we keep $w=1/N$, with $N\rightarrow\infty$, $\lambda_{3,4}\in\mathbb{C}$ with $\text{Re }\lambda_{3,4}<0$ so that $S_1$ and $S_2$ are saddle-foci. The last fixed point $S_{12}$, which lies on the diagonal plane $D$, has two eigenvalues $\lambda_1>0$ and $\lambda_2<0$, as well. The other two read $\lambda_{3,4} = \frac{1}{2} (-\xi \pm \sqrt{\xi^2 + 4 \sigma (1-4w) \cos\alpha / R''^2})$. With $N\rightarrow\infty$, we have $\lambda_3>0$ and $\lambda_4<0$. Therefore, $S_{12}$ is a saddle with two stable and two unstable manifolds.

The homoclinic bifurcation discussed previously leads to the creation of two periodic orbits $\gamma_1$ and $\gamma_2$ which lie in the respective two-dimensional subspaces
\begin{equation}
\label{eq:subspace_of_first_solitary_particle}
	P_1 := \left\{ (\Delta\varphi_1,\Delta\omega_1,\Delta\varphi_2,\Delta\omega_2) \in \Omega^2 \mid \Delta\varphi_2=0, \Delta\omega_2=0 \right\}
\end{equation}
and $P_2 := \left\{ (\Delta\varphi_1,\Delta\omega_1,\Delta\varphi_2,\Delta\omega_2) \in \Omega^2 \mid \Delta\varphi_1=0, \Delta\omega_1=0 \right\}$. The phase portrait in $P_1$ is shown in Fig.~\ref{fig:projected_phase_portrait_for_two_solitary_particles}(a) (in $P_2$, it is similar due the symmetry of \eqref{eq:ode_for_two_differences_symmetric}). At this point, all trajectories on the diagonal plane $D$, where the saddle point $S_{12}$ exists, converge to the focus $O$ (except for the saddle itself and its stable manifolds) (cf. Fig.~\ref{fig:projected_phase_portrait_for_two_solitary_particles}(b)). 
With a subsequent increase of the phase lag till some $\alpha=\alpha_2$, the next homoclinic orbit $\gamma_{12}$ is created, belonging to the diagonal plane $D$. It occurs in the moment when the unstable manifold of the saddle $S_{12}$ merges with its stable manifold. With $\alpha>\alpha_2$, the phase portrait on the diagonal plane contains two equilibria $O$ and $S_{12}$ and a limit cycle (cf. Fig.~\ref{fig:projected_phase_portrait_for_two_solitary_particles}(d)) that signifies existence of two solitary oscillators, which rotate in-phase and with the same frequency.

\begin{figure*}[]
	\centering
	\includegraphics[width=1.0\textwidth]{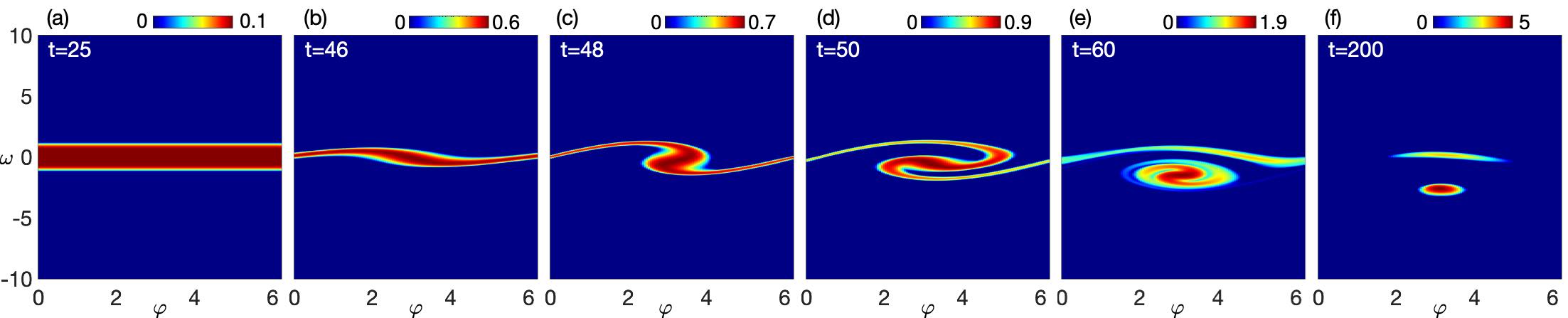}
	\caption{Temporal evolution of a PDF as a solution to the continuum Kuramoto model with inertia \eqref{eq:continuum_limit_pde} towards a solitary state with one frequency cluster starting from irregular initial conditions (see SM and \cite{bcs_youtube_channel} for a corresponding movie). Snapshots are taken at (a) $t=25$, (b) $t=46$, (c) $t=48$, (d) $t=50$, (e) $t=60$, and (f) $t=200$. Color denotes values of the PDF. The discretization in $(\varphi,\omega)$ is $128\times1000$. Other parameters are $\xi=0.1$, $\sigma=1.0$, $\alpha=0.3$, $D_\varphi=0.001$.}
	\label{fig:temporal_evolution_of_probability_density_function}
\end{figure*}

With a subsequent increase of $\alpha$, we observe a cascade of homoclinic bifurcations appearing on diagonals defined by $(\Delta\varphi_1,\Delta\omega_1) = (\Delta\varphi_2,\Delta\omega_2)$, $(\Delta\varphi_1,\Delta\omega_1) = (\Delta\varphi_2,\Delta\omega_2) = (\Delta\varphi_3,\Delta\omega_3)$, and so on.
This behavior depends on the fraction $w$ of solitary oscillators. Therefore, with the increase of the population size, the maximum allowed size of a solitary group increases linearly. However, in the large population size limit $N\rightarrow\infty$, the probability to observe any finite \textit{number} of solitary oscillators out of the whole population tends to zero and we become interested in observing a finite \textit{fraction} of solitary oscillators, which will consist of an infinite subpopulation. In this limit, since $w\rightarrow0_+$, all homoclinic bifurcation boundaries, described above, collapse into one that signifies the appearance of the whole solitary group (cf. Fig.~\ref{fig:solitary_cascade}(a)).




\section{Langevin dynamics}

In the context of active matter and in general in systems where global polar order emerges, it is convenient to introduce a respective order parameter that characterizes large scale oscillatory ensembles in low dimension. The classical macroscopic characteristic for such systems is a global polar order parameter defined as a complex valued function as
\begin{equation}
\label{eq:polar_order_parameter}
	R(t)e^{i\Theta(t)} = \frac{1}{N} \sum_{j=1}^{N} e^{i\varphi_j(t)}.
\end{equation}
If we consider phases as orientations of self-propelled particles, this operation is equivalent to computing the average direction of motion of all population. The magnitude shows how strongly the system is synchronized. If $R=0$, it is disordered, and if $R=1$, it is completely synchronized. Otherwise, $R\in(0,1)$ indicates a partial level of orientational order. Note that one can rewrite the interaction term in \eqref{eq:particle_ode_global} in terms of this order parameter, i.e., $\frac{1}{N} \sum_{j=1}^N \sin(\varphi_j - \varphi_i - \alpha) = R \sin(\Theta - \varphi_i - \alpha), \,i=1,\dots,N$. Namely, the dynamics of one oscillator are defined in terms of a mean field. This mean field is in turn generated by the whole population of oscillators. This formulation will prove useful for our subsequent study of the mean-field limit behavior.



From the natural and application point of view, we must take into account that oscillators are generally subject to some external perturbations which we consider to be of stochastic origins. Therefore, we now regard oscillators as interacting stochastic processes and reformulate ODEs \eqref{eq:particle_ode_global} as stochastic differential equations (SDEs)
\begin{equation}
\label{eq:particle_sde1}
\begin{aligned}
	\mathrm{d}\varphi_i(t) &= \omega_i(t) \;\mathrm{d}t, \\
	\mathrm{d}\omega_i(t) &= -\xi\omega_i(t) \;\mathrm{d}t + \sigma R(t) \sin(\Theta(t) - \varphi_i(t) - \alpha) \;\mathrm{d}t \\
	&+ \sqrt{2D_\varphi} \;\mathrm{d}W_i(t),
\end{aligned}
\end{equation}
where interactions between oscillators are expressed in terms of the mean field as previously discussed. Oscillators are subject to external stochastic forces accounted for by families of independent Wiener processes $(W_i(t))_{t\geq0}, \,i=1,\dots,N$ with $D_\varphi>0$ as the noise strength.
The resulting dynamics under noise are determined through the interplay of alignment and stochastic forces. Namely, if the first one prevails, we observe the emergence of ordered motion; otherwise, the motion remains disordered.

The temporal evolution of a network of oscillators towards a solitary state with one frequency cluster under noise is qualitatively similar as in a deterministic setup (cf. Fig.~\ref{fig:space_time_plots_sorted}). Starting from random initial conditions (with compact support for frequencies), oscillators synchronize in frequency first with phases remaining disordered. At some later point, oscillators start to synchronize in phases. However, each frequency cluster is no longer characterized with a single averaged value. Instead, each cluster is now characterized with a distribution over $\omega$. For partial synchronization, this distribution is skewed and unimodal. For solitary states, it is multimodal, where the number of peaks corresponds to the number of frequency clusters with the highest peak representing the largest synchronized group. For small $D_\varphi$, averaged frequencies of all groups are less than zero (for $\alpha>0$). With the increase of $D_\varphi$, they shift towards zero until all modes coalesce and the distribution for the whole population becomes Gaussian for $\omega$ and uniform for $\varphi$ (compare to Eq.~\eqref{eq:gaussian_density_function}).



\section{Mean-field limit}

To understand the dynamics of solitary states without finite size effects, we turn to the mean-field description of interacting particle systems, i.e., the limit $N\rightarrow\infty$ \cite{canizo2011mmmas,carrillo2014jsp,lancellotti2005ttsp,neunzert1984springer,kipnis1998scaling}. This description is commonly provided by a one-particle probability density function (PDF) $f(\varphi,\omega,t) : \Omega\times\mathbb{R}_+ \rightarrow \mathbb{R}_+$ with $\Omega = \mathbb{T} \times \mathbb{R}$, which quantifies the probability to find an oscillator having phase $\varphi$ and rotating with frequency $\omega$ at time $t$. Our first goal in this section is to find a partial differential equation (PDE) that governs its evolution.
We start by considering an empirical PDF
$
	f^{[N]}(\varphi,\omega,t) = \frac{1}{N}\sum_{i=1}^{N} \delta[\varphi - \varphi_i(t)] \delta[\omega - \omega_i(t)]
$
as a particle approximation to the mean-field PDF. This function gives the fraction of oscillators that have phase $\varphi$ and frequency $\omega$ at time $t$. Using the framework of Fokker-Planck equations \cite{risken}, we look for an ensemble-averaged representation of the empirical PDF \cite{archer:jpa}. This way, one obtains an infinite hierarchy of $n$-particle density functions and assumes a mean-field approximation $f(\varphi_1,\omega_1,\varphi_2,\omega_2,t) \approx f(\varphi_1,\omega_1,t) f(\varphi_2,\omega_2,t)$ in order to close the hierarchy at the first order. As a result, we obtain a nonlinear Fokker-Planck-Kolmogorov PDE for a one-particle PDF $f=f(\varphi,\omega,t)$
\begin{equation}
\label{eq:continuum_limit_pde}
\begin{aligned}
	\partial_t f &= -\omega \partial_\varphi f \\
	&- \partial_\omega\left\{ f \left[ -\xi\omega + \sigma R \sin(\Theta - \varphi - \alpha) \right] \right\} + D_\varphi\partial_{\omega\omega} f,
\end{aligned}
\end{equation}
where the polar order parameter \eqref{eq:polar_order_parameter} becomes
\begin{equation}
\label{eq:polar_order_parameter_in_continuum_limit}
	R(t)e^{i\Theta(t)} = \int_\Omega e^{i\varphi} f(\varphi,\omega,t) \;\mathrm{d}\varphi\mathrm{d}\omega.
\end{equation}

\begin{figure}[]
	\centering
	\includegraphics[width=0.5\textwidth]{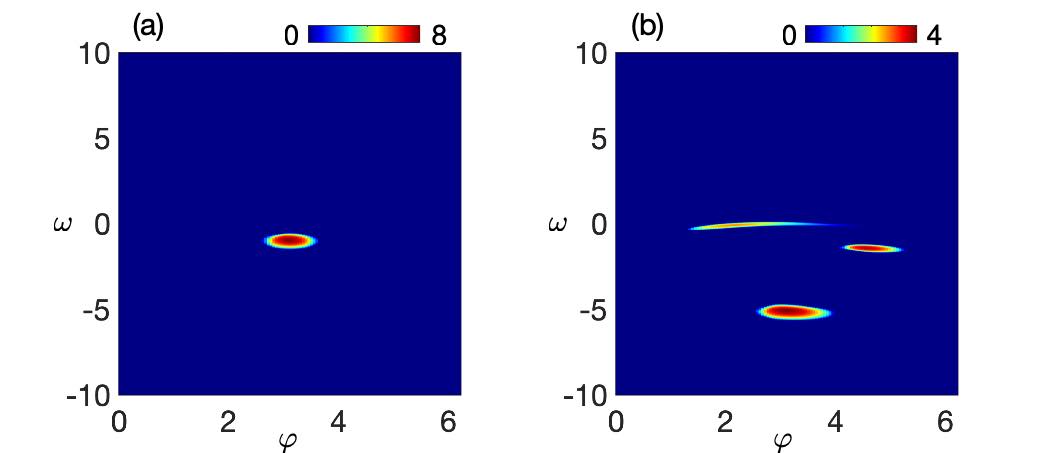}
	\caption{Snapshots of solutions of the continuum Kuramoto model with inertia Eq.~\eqref{eq:continuum_limit_pde} representing (a) nonstationary partial synchronization ($\alpha=0.1$, $D_\varphi=0.001$) and (b) a solitary state with two frequency clusters ($\alpha=0.8$, $D_\varphi=0.0001$) (see SM and \cite{bcs_youtube_channel} for corresponding movies). Color denotes values of a density function $f(\varphi,\omega,t)$. Grid discretization in $(\varphi,\omega)$ is $128$ by $1000$. Other parameters are $\xi=0.1$, $\sigma=1.0$.}
	\label{fig:continuum_limit}
\end{figure}

An example of how the mean-field dynamics evolves towards a solitary state with one frequency cluster is presented in Fig.~\ref{fig:temporal_evolution_of_probability_density_function} (see SM and \cite{bcs_youtube_channel} for a corresponding movie). One observes qualitatively similar evolution of a PDF $f$ and a finite-size oscillator ensemble represented in terms of a coarse-grained PDF (cf. Fig.~\ref{fig:temporal_evolution_of_probability_density_function} and Appendix~D). Namely, starting from a sufficiently irregular initial condition with compact support, the solution rapidly becomes uniform in $\varphi$ and unimodal in $\omega$. This signifies the tendency of oscillators to synchronize via frequency (cf. Fig.~\ref{fig:temporal_evolution_of_probability_density_function}(a)). When such synchronization is large enough, the high-density stripe starts to bend and rotate (cf. Fig.~\ref{fig:temporal_evolution_of_probability_density_function}(b-d)). This means that oscillators begin to synchronize in phase, which is updated via frequency. Therefore, the probability mass that lies above the $\omega=0$ line, warps one way while the lower part warps the opposite way. The direction of rotation depends on the sign of $\alpha$. Such rotational motion is also contracting. If $\alpha$ is large enough, a part of probability mass separates and follows its own oscillating trajectory (cf. Fig.~\ref{fig:temporal_evolution_of_probability_density_function}(e)). Eventually, both peaks become more compact (cf. Fig.~\ref{fig:temporal_evolution_of_probability_density_function}(f)), the level of which depends on a diffusion constant $D_\varphi$.
By varying $\alpha$, one observes partial synchronization or a solitary state with two frequency clusters, respectively (cf. Fig.~\ref{fig:continuum_limit}). For all numerical studies of \eqref{eq:continuum_limit_pde}, we have implemented a finite volume method \cite{kruk2020fvm} with boundary conditions being periodic for $\varphi$ and zero flux for $\omega$.

\begin{figure}[b]
	\centering
	\includegraphics[width=0.5\textwidth]{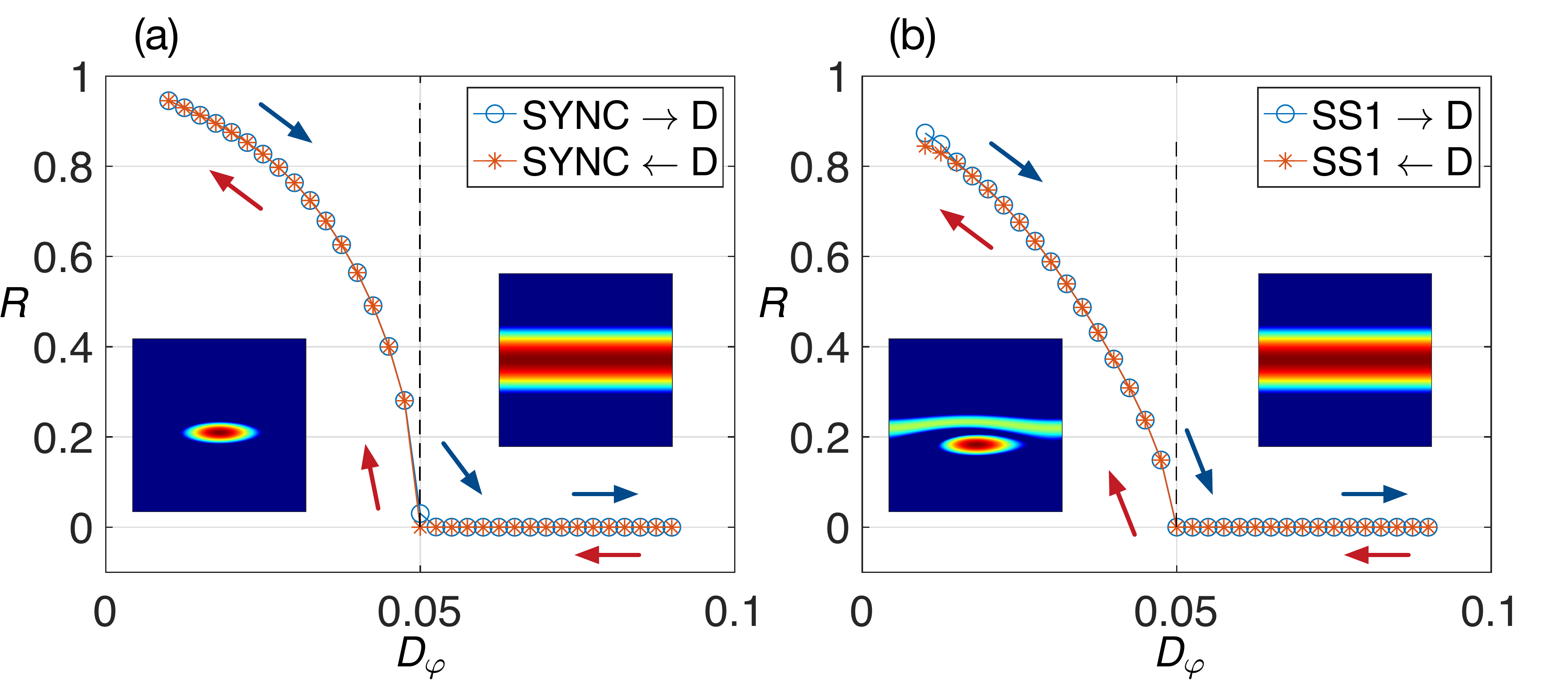}
	\caption{Phase transitions between disordered motion (D) and (a) partial synchronization (SYNC) and (b) a solitary state with one frequency cluster (SS1). The transitions are quantified in terms of a polar order parameter magnitude $R$ Eq.~\eqref{eq:polar_order_parameter_in_continuum_limit}. Routes from SYNC and SS1 towards D are depicted with blue circles. Routes from D to SYNC and SS1 are depicted with red stars. Black dashed lines denote order-disorder transition points $D_\varphi=\frac{1}{2}\sigma\xi\cos(\frac{1}{2}\alpha)$ Eq.~\eqref{eq:order_disorder_transition_line_nonzero_lag}. Insets show PDFs at respective endpoints of bifurcation curves. Colored arrows indicate directions of bifurcation paths. Other parameters are $\xi=0.1$, $\sigma=1$, (a) $\alpha=0.1$, (b) $\alpha=0.3$.}
	\label{fig:phase_transitions}
\end{figure}

From the stochastic dynamics \eqref{eq:particle_sde1}, we know that oscillators can synchronize in one group or in several groups, i.e., a solitary state, or remain disordered, i.e., $R=0$. The latter case is described with a stationary distribution in the mean-field limit. To find its analytical expression, we first note that due to periodic boundaries, the PDF in terms of $\varphi$ is uniform. Therefore, it should only depend on frequency $\omega$, over which it has unbounded support. Second, we require that the PDF is sufficiently regular, i.e., $f(\omega)\rightarrow0$ and $f'(\omega)\rightarrow0$ as $\omega\rightarrow\pm\infty$. Under such assumptions, one can show that disordered oscillatory motion in the mean-field limit is described by (cf. Fig.~\ref{fig:phase_transitions}, the right insets)
\begin{equation}
\label{eq:disordered_density_function}
	f(\varphi,\omega) = \frac{1}{2\pi} \Phi_G(\omega),
\end{equation}
where a prefactor $1/(2\pi)$ arises due to normalization with respect to $\varphi$ and $\Phi_G$ is a Gaussian PDF
\begin{equation}
\label{eq:gaussian_density_function}
	\Phi_G(\omega) = \frac{1}{\sqrt{2\pi s^2}} e^{-\frac{\omega^2}{2 s^2}}
\end{equation}
with zero mean and standard deviation $s = \sqrt{D_\varphi / \xi}$. Note that this solution is valid for any phase lag value $\alpha$.

In the absence of the phase lag and for sufficiently small noise, oscillators exhibit partial synchronization. A PDF that corresponds to this behavior is symmetric with respect to both $\varphi$ and $\omega$. In fact, the $\omega$-marginal has qualitatively the same form as Eq.~\eqref{eq:gaussian_density_function}. Therefore, we look for a stationary solution of the form $f(\varphi,\omega) = \Phi(\varphi)\Phi_G(\omega)$, where $\Phi$ is to be determined. Substituting $f$ into Eq.~\eqref{eq:continuum_limit_pde}, we find
\begin{equation}
\label{eq:synchronized_density_function}
	f(\varphi,\omega) = \Phi_{VM}(\varphi) \Phi_G(\omega),
\end{equation}
where $\Phi_{VM}$ is a von Mises PDF
\begin{equation*}
	\Phi_{VM}(\varphi) = \frac{1}{2\pi I_0(\gamma)} e^{\gamma \cos(\varphi - \Theta)}
\end{equation*}
with $\gamma = \sigma \xi R / D_\varphi$ denoting system's relative synchronization level and $I_0$ is the modified Bessel function of the first kind. This PDF depends on $R$ which in turn depends on the PDF itself. Thereby, we can determine $R$ implicitly from
$
	R = \dfrac{I_1[\gamma(R)]}{I_0[\gamma(R)]}
$
and put $\Theta\equiv0$ without loss of generality due to the translation invariance of \eqref{eq:continuum_limit_pde}. Analyzing this expression around the onset of orientational order at $R=0$ \cite{kruk2020pre}, we find the order-disorder transition line
\begin{equation}
\label{eq:order_disorder_transition_line_zero_lag}
	D_\varphi = \frac{\sigma \xi}{2}.
\end{equation}
For noise levels higher than this critical value, oscillatory motion remains disordered, i.e., $R=0$ for all $t$, while for lower values of $D_\varphi$, one observes the emergence of polar order described by \eqref{eq:synchronized_density_function}. The transition across this line is of second order and it corresponds to supercritical pitchfork bifurcation in terms of $R$. This is reminiscent of the continuum Kuramoto model with noise \cite{carrillo2019jcp}.

For the more interesting collective motion of polar order with rotation, i.e., $\alpha>0$, and solitary states, we cannot find analytical solutions to \eqref{eq:continuum_limit_pde}. These are composed of one or several skewed bivariate peaks (cf. Figs.~\ref{fig:temporal_evolution_of_probability_density_function} and \ref{fig:continuum_limit}). But since we know the representation of disordered motion for any $\alpha$, we perform linear stability analysis of it against perturbations in Fourier space (see the details in Appendix~E). As a result, we obtain stability diagrams for \eqref{eq:disordered_density_function} in the parameter domain of $D_\varphi$ and $\alpha$. We find that the transition line where this solution becomes unstable is well described by
\begin{equation}
\label{eq:order_disorder_transition_line_nonzero_lag}
	D_\varphi = \frac{\sigma \xi}{2}  \cos\left( \frac{\alpha}{2} \right),
\end{equation}
which is consistent with \eqref{eq:order_disorder_transition_line_zero_lag}. In particular, we see that in the deterministic case $D_\varphi\rightarrow0_+$, oscillators never stay disordered, as mentioned before. The knowledge of the general order-disorder transition line \eqref{eq:order_disorder_transition_line_nonzero_lag} allows us to study phase transitions between disordered motion and partial synchronization or solitary states. We find that in both cases, the transition is of second order (cf. Fig.~\ref{fig:phase_transitions}) which is similar to the results for the continuum Kuramoto model with noise for identical oscillators \cite{carrillo2019jcp,kruk2020fvm}. We note that for the continuum Kuramoto model with inertia for nonidentical oscillators (cf. Appendix~F) and in the absence of the phase lag \cite{tanaka1997prl,tanaka1997physica_d,olmi2014pre,munyaev2020njp} first order transitions accompanied by hysteresis effects were reported.

\section{Conclusions}

In this paper, we have considered self-propelled particle systems where the direction of motion of a particle obeys a second-order differential equation. We have limited our attention to spatially homogeneous configurations of particle ensembles which allowed us to simplify their equations of motion. This way, we have come to the Kuramoto model with inertia. This model is particularly known for a phenomenon of solitary states. Our interest is to understand how they would manifest themselves in the light of active matter theory. Our present analysis therefore facilitates a subsequent study of respective particle dynamics in a spatially inhomogeneous setup. 

We have described the emergent oscillatory dynamics with a special emphasis on the large scale limit. We have found that in addition to (partial) synchronization, one observes solitary states with potentially arbitrary number of frequency clusters. This result holds in both deterministic and stochastic setups. Our mean-field formulation of oscillatory dynamics demonstrates the existence of solitary states with different number of frequency clusters as well. Moreover, we have established that the phase transition between polar order and disorder for the continuum Kuramoto model with inertia for identical oscillators is of second order which is in contrast to the nowadays reported cases where oscillators have different natural frequencies. However, the extension of the present work to the case of nonidentical oscillators would be subject to future work.

\section*{Supplementary material}

See supplementary material for the derivation and nonlinear dynamical analysis of difference equations for one and two solitary particles, the linear stability analysis of the mean-field PDE, and accompanying movies.

\begin{acknowledgments}
HK acknowledges support from the European Research Council (ERC) with the consolidator grant CONSYN (grant no. 773196).
\end{acknowledgments}

\section{Data availability}

The data that supports the findings of this study are available within the article [and its supplementary material].

\bibliography{solitary_states_in_the_mean_field_limit}

\end{document}